# Technostress and Resistance to Change in Maritime Digital Transformation: A Focused Review

Benedicte Frederikke Rex Fleron, Roskilde University, Roskilde, Denmark

Raluca Alexandra Stana, Roskilde University, Roskilde, Denmark

**Abstract.** The maritime industry is undergoing a significant digital transformation (DT) to enhance efficiency and sustainability. This focused review investigates the current state of literature on technostress and resistance to change among seafarers as they adapt to new digital technologies. By critically reviewing a focused selection of peer-reviewed articles, we identify the main themes and trends within maritime research on DT. Findings indicate that while mental health issues are a predominant concern, this is yet to also be investigated in the context of new technology introduction in an industry that is already setting seafarers under pressure. Additionally, change management is not addressed, and DT is limited to specific functionalities rather than embracing broad work practice transformations.

## 1. Introduction

The maritime industry, pivotal to global trade, is undergoing a significant digital transformation (DT) (Jović et al., 2022). This transformation encompasses the integration of advanced digital technologies into maritime operations, aiming to enhance efficiency, safety, and environmental sustainability (World Maritime University et al., 2023). As the sector adopts technologies such as automation, data analytics, and interconnected systems, the implications of these changes extend beyond operational workflows to affect the human elements of maritime work (Cichosz et al., 2020; Karlsson et al., 2023).

Incorporating human factors into the discourse digital transformation is generally crucial, as overwhelming information systems research suggests (Riedl et al., 2012; Sarker et al., 2019; Wibowo et al., 2022), thus it is imperative for the maritime sector to also incorporate this in both research and practice. As the maritime industry evolves, understanding the human implications of these technologies—how they affect crew operations, job satisfaction, and overall well-being—is vital. Neglecting these aspects could hinder successful DT implementation, potentially exacerbating challenges like technostress (World Maritime University et al., 2023), job dissatisfaction, and resistance to change among seafarers (Cichosz et al., 2020; Karlsson et al., 2023).

Thus, there is still an opportunity to understand what the current state of research within the maritime digital transformation is. This paper seeks to address the question: *What are the current gaps and trends in understanding technostress and resistance to change within the context of maritime digital transformation?*

Methodologically, we will answer our research question through a focused literature review (Grant & Booth, 2009; Snyder, 2019), in order to identify how some of the core challenges connected to technostress and resistance to change in achieving successful DT are currently addressed in maritime digital transformation research.

By doing this, we hope to highlight what is yet to be explored within research on maritime digital transformation and propose a future paths of research agenda.

# 2. Theoretical Background

Seafarers' unique lifestyle and challenging working conditions on vessels distinctly impact their family life, mental, physical, and spiritual well-being due to prolonged absences from home and high-risk environments (Dewan & Godina, 2023). Abucay (2023) highlights the need for integrating care for mental, physical, and spiritual health to support seafarers' overall well-being. Additionally, Hui (2019) suggests physical exercises to mitigate health issues like depression, exacerbated by factors such as family separation, intense work hours, and safety risks. The increasing use of digital technologies further intensifies work pressures (World Maritime University et al., 2023).

## 2.1 Digitalization and Digital Transformation

There is a profound difference between digitalization and digital transformation. While transformation is not a new concept, it has become increasingly relevant in our daily lives, particularly concerning green transitions.

Digitalization involves several processes. Primarily, it refers to converting analogue data into digital form, transforming analogue information into binary numbers (bits). This process is essential for computers to understand and manipulate data. Digitalization also encompasses a sociotechnical dimension where digital techniques are employed to transform organizations, business models, and practices. The distinction lies between the technical process of digitizing data and the broader sociotechnical processes where these digital techniques create new digital artifacts, enabling companies or organizations to fundamentally change their business models and practices (Brennen & Kreiss, 2016; Schallmo & Williams, 2018; Tilson et al., 2010).

In the maritime research and practice field, digitalization primarily concerns the automation of inefficient processes and work practices, along with the conversion of analogue artifacts and data into digital form. The implementation and use of digital technologies increase work pressure on seafarers, especially when they must interact with multiple systems and face a lack of skills and education in operating these technologies. Therefore, it is crucial for management to be aware of the speed at which such technologies are implemented, the interoperability of these systems with existing ones, and the seafarers' capabilities in operating these technologies, considering the known strains of seafarer work.

Digital transformation, although not confined to a single definition, has been described in the Information Systems literature by Vial (2019) as "a process that aims to improve an entity by triggering significant changes to its properties through combinations of information, computing, communication, and connectivity technologies." These technologies include social, mobile, analytic, cloud, Internet of Things (IoT), platforms, and ecosystems. This process of digital transformation is yet to be fully realized within the maritime sector, highlighting the need for significant transformation in this industry.

The increased work pressure, the need for new skills, and the rapid pace of technological change all contribute to resistance and technostress among employees. Understanding and addressing these factors are critical for successful digital transformation.

## 2.1. 2.2. Resistance to Change

Significant organizational change is often closely followed by resistance. This common phenomenon reflects the challenges that organizations face when introducing new technologies or processes, as these changes can disrupt established workflows and employee routines. Understanding this resistance is crucial for effectively managing the transition and ensuring successful implementation of digital strategies.

There is no single definition of resistance to change as it is addressed from various perspectives across different fields. Within the project management literature, resistance is often examined at the theme or social group level. One classic approach to understanding resistance is through Kurt Lewin's concept of force field analysis, which identifies the forces for and against change. This method, dating back to the 1940s, is utilized in works like that of Coch & French (1948) to formulate resistance to change theories.

In project management, the work often focuses on five key concepts: the difficulty of the work, the goal to be achieved, management pressure, group standards, and social pressures. The social group pressure is typically the most influential force, setting standards for individuals in terms of resistance or acceptance of change. Effective communication and support of social groups are essential strategies for management to mitigate potential resistance. Engaging affected employees in the change process is crucial.

From a psychological perspective, resistance is often viewed as the individual's negative experience towards the change process. Oreg (2003) discusses this in terms of regression analysis, where individuals are asked about their perceptions of the impending change, including the aim, process, information, and their role in it. Within the information systems research field, resistance is frequently linked to knowledge or the lack thereof, especially concerning technology or IT skills required for new tasks.

For nearly 50 years, the view that resistance is an individual or group-related concept, with management needing to remove or alleviate it, remained largely unchallenged. However, Dent & Goldberg (1999) argued that people resist the loss of status, pay, or comfort rather than change itself. This resistance is more about the concern for the impact of change, such as its effect on social status, livelihood, and social relations, rather than the change itself.

Classical approaches to change include Kurt Lewin's change model, which involves the steps of unfreezing, changing, and refreezing the organization. This linear model is often challenged by the rapid pace of technological development, which requires continuous adaptation rather than static stages. Kotter's (1995) eight steps to organizational change emphasize creating a sense of urgency and a reason for change, which helps employees understand and engage with the change process.

Lapointe & Rivard (2005) view resistance to IT as a potential power-shifting medium within organizations. The use of technology can alter power dynamics among social groups and between organizations. This shift can be generational or educational, with younger or more educated employees better utilizing technology. Resistance, therefore, becomes an obstacle to be removed for successful IT system implementation.

Bagayogo et al. (2013) argue against treating resistance at face value, advocating for understanding the root concerns, which may include quality issues or risks associated with new systems. Resistance can sometimes have a positive impact by highlighting potential flaws or risks.

Shimoni (2017) identified three schools of thought on resistance to change: the individual psychological disposition perspective, the social context perspective, and the social construction perspective. He also proposed a fourth perspective, the habitus-oriented approach based on Bourdieu's action theory, which views resistance as an internalized and institutionalized part of an organization's social field.

Smollan (2011) highlight the multidimensional nature of resistance, which includes cognitive, affective, and intentional components. Resistance can be expressed consciously or unconsciously, adding complexity to its understanding. Researchers and practitioners must address both observable and less detectable aspects of resistance, including emotional and affective responses.

Resent work by (Fleron et al., 2023) presents a framework for encountering resistance, addressing both outspoken and silent, observable and non-observable, deliberate and non-deliberate actions of individuals and groups. The framework considers who is resisting, what is being resisted (the process, change itself, or technology), and the specific stakeholders involved. It also incorporates a temporal aspect, focusing on past, present, and future resistance.

This framework aims to better understand and manage resistance to change in information systems-related projects, ultimately leading to more successful change initiatives.

### 2.2. 2.3. Technostress, the dark side of digitalization

Technostress, or the direct and indirect stress experienced from interactions with digital technologies, was first coined in 1983 by Craig Brod (Brod, 1984). Since then, the concept of technostress has permeated various disciplines including library studies, computer science, and information studies. A prevalent model for researching technostress is the transactional model, which is deeply rooted in stress research, and which sees (techno)stress as a dynamic and processual interaction between stressors and strains (Cooper et al., 2001; Lazarus & Folkman, 20). This model primarily studies the root causes of technostress—known as stressors—and the impact of these stressors, or strains. Additionally, it considers coping strategies, appraisal processes, antecedents, and outcomes, as well as factors that inhibit technostress (La Torre et al., 2019; Nisafani et al., 2020; Sarabadani et al., 2018).

Despite the term being established in 1983, it is only in recent years that technostress has gained significant attention and become more thoroughly investigated. Among the antecedents—factors that influence how one experiences technostress—are age (Hauk et al., 2019), gender (Ma & Turel, 2019), culture (Tu et al., 2005), and computer literacy (Tarafdar et al., 2011).

Established literature identifies several well-recognized stressors that contribute to technostress, including technical complexity(Tarafdar et al., 2007), frequent computer upgrades, usability errors(Ayyagari et al., 2011), utility issues (Hertzum & Hornbæk, 2023), system performance, and terminology mismatches(Califf et al., 2020).

Stressors can vary in their impact on individuals, ranging from mild to severe strains. Examples include headaches, fatigue, back problems, and eye strain. More broadly, stress has also been linked to anxiety, depression, and in extreme cases, life-threatening conditions such as cancer (Jensen et al., 2017). Recent research in digitalization and technostress has found that organizations with higher levels of digitalization report that their employees experience elevated levels of stress (Gimpel et al., 2022).

Given the maritime industry's desire to digitally transform their business models and, consequently, digitalize work processes and operations, it is crucial for the sector to consider the implications of technostress. This consideration is essential to safeguard the well-being of seafarers as they navigate these changes.

# 3. Methodology

We conducted a focused literature review aimed at understanding how digital technologies impact maritime workers, specifically exploring technostress, job satisfaction, well-being, and resistance to change and technology non-use. Given the sparse direct research on technostress within this sector, we expanded our scope to include related aspects of job satisfaction and well-being.

### 3.1. Data Collection

We selected databases with broad coverage of scientific publications, including "Scopus", "Web of Science", "Business Source Complete", and the electronic library of AIS (Kabra et al., 2024). Our search was limited to peer-reviewed articles published from 2013 to present. The search criteria included terms related to mental health, technostress, job satisfaction, resistance to technology use, and digital transformation.

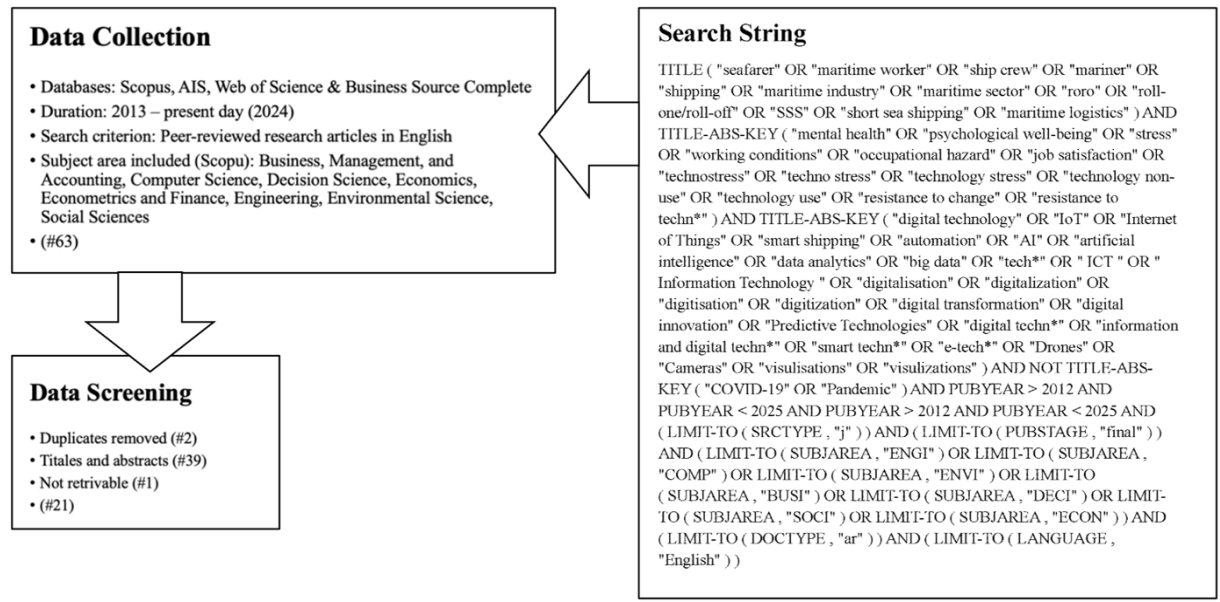

*Figure 1: Exhibit of the data handling process*

### 3.2. Screening Process

The initial search provided a diverse set of articles from Scopus (25), AIS (4), Web of Science (4), and Business Source Complete (28). After removing duplicates, titles and abstracts were screened for relevance. Articles not directly addressing the research questions were excluded. This was followed by a thorough review of the full texts to ensure relevance to our study aims. The process resulted in a final selection of 21 articles, with content spanning mental health (8), job satisfaction (8), digital technology impacts (10), and barriers to digital transformation (8). The specifics of our data handling process are detailed in Figure 1: Exhibit of the data handling process.

## 4. Findings

Our categorization of papers reveals that most studies predominantly employ quantitative methods, with some exceptions engaging in meta-reviews (Hui, 2019), neurobiological measurements (Liu et al., 2020), qualitative (Ghaderi, 2019; Kongsvik et al., 2014; Raza et al., 2023), mixed-methods (Abucay, 2023) or action research (Thiess & Müller, 2018). Considering the unique maritime contexts within which technology is deployed, future research might focus on understanding the subjective experiences of seafarers through qualitative research.

Additionally, the papers underscore a significant focus on seafarers' mental health: burnout (Turan et al., 2022), fatigue (Bal BeşİkÇİ et al., 2016; Bal et al., 2015), occupational diseases (Hui, 2019), or examine counterproductive work behaviours (Uche et al., 2018), seafarers' quality of life onboard (Abucay, 2023; Kaya & Yorulmaz, 2023), or intentions to stay (Yao & Huang, 2019), typically at a macro-industry level, though not connecting these issues directly to DT. However, given the maritime sector's fragmentation and the variation in job types, both onshore and offshore, it is crucial to consider this diversity when conducting research related to stress and burnout. Future research might consider both on- and off-shore workers, their various cultural backgrounds, and how DT might be associated with mental health issues, including technostress.

In terms of technology, while one study investigates the micro-level impacts of digitalization and automation on seafarers' work processes (Li et al., 2022), and others explore mandatory technology usage and its link to technology withdrawal (Daud et al., 2022) and autonomous technology (Ghaderi, 2019), these investigations remain fragmented. There is scant integration between digital transformation and maritime operations, with only one study directly connecting digital transformation with seafarer concerns (Raza et al., 2023).

Currently, no studies explicitly connect resistance to change with technostress in maritime settings, nor do they adequately address the direct impacts of technostress on seafarers. This lack of focus presents a critical opportunity for future research to explore how technostress affects seafarers navigating digital transformation challenges. Moreover, recent research fails to link mental health issues such as burnout and fatigue among seafarers to technostress and technology use (Abucay, 2023; Bal et al., 2015; Hui, 2019). This disconnect suggests an industry-wide underestimation of technostress's effects, risking the well-being of employees amid digital shifts. Given the high risk of stress-related incidents among seafarers, it is vital that digital transformation strategies proactively address and mitigate technostress.

## 5. Discussion

The literature reveals a shallow examination of how digital technologies impact mental health within the shipping industry. Automation and digital transformation lead to reduced crew sizes, increasing the workload for the remaining staff, and resulting in extended working hours and diminished support among crew members, which exacerbates mental health challenges (Ghaderi, 2019). To counterbalance these issues, there needs to be a balance between technological advancements and the well-being of crew members. As operations become more digitised, the demand for crew members to develop new skills in technology and data analysis grows, as it would in other industries as well (Wibowo et al., 2022).

Education and supportive systems are essential to help crew adapt to these new demands and mitigate the increased mental strain. This scenario mirrors wider industry trends where technological efficiencies often lead to reduced staff numbers, underlining a common misconception that technology should replace rather than augment human capabilities and enhance overall value. The industry needs to recognize the shift in skill sets and identity required for today's seafarers, promoting the development of flexible, adaptable capabilities suitable for the diverse configurations of modern vessels.

The literature review indicates that resistance to change within the maritime sector is underexplored, focusing instead on mental health, risk assessment, optimization, and autonomy (Bal BeŞİkÇİ et al., 2016; Bal et al., 2015; Ghaderi, 2019; Li et al., 2022; Turan et al., 2022). There seems to be minimal understanding of the existence or nature of this resistance, raising questions about its roots—whether it stems from poor change management, inherent opposition to change, or specific resistance to new technologies (Fleron et al., 2023). For example, workarounds or the incorrect use of technology might reflect concerns over maintaining work quality rather than outright resistance or skill deficits (Fleron et al., 2023).

Furthermore, the review reveals a limited number of articles addressing comprehensive digital transformation, with the existing literature often limiting its scope to digitization or digitalization that enhances specific functionalities rather than broad work practices (Agatić & Kolanović, 2020; Brouer et al., 2018; Ghaderi, 2019; Raza et al., 2023; Thiess & Müller, 2018). This narrow focus suggests that the maritime industry's engagement with digital technologies typically does not advance towards a full digital transformation but remains at enhancing operational aspects, thus not fully capturing the potential benefits of digital technologies (Parola et al., 2021). This gap emphasizes the need for future studies to examine both the efficacy of technologies and their broader psychological and sociological impacts (Agatić & Kolanović, 2020; Parola et al., 2021).

## 6. Conclusion

This review critically examines DT in the maritime industry, focusing on technostress and resistance to change among seafarers. It highlights the necessity of integrating human factors into DT initiatives, as technological advancements like automation significantly impact seafarer well-being and job satisfaction.

The implications for research are clear: there is a crucial need for more empirical studies focusing on the human aspects of maritime DT. Future research should explore the full impact of DT on mental health and job satisfaction to better equip the maritime industry to address these challenges effectively, ensuring a holistic and sustainable approach to digital transformation.